\RequirePackage{fix-cm}
\documentclass[11pt,a4paper]{article}      

\usepackage{graphicx}
\usepackage{latexsym,amssymb}
\usepackage[centertags]{amsmath}
\usepackage{epsfig,pifont,rotating}
\usepackage[inactive]{srcltx}  

\usepackage{graphics,helvet,times,mathptm,epic,eepic}

\usepackage{color}

\usepackage{newlfont}

\usepackage{amsfonts}

 \newenvironment{keywords}{ \noindent {\small\bf Key Words}:}{ }

\def\bd{\begin{description}}
\def\ed{\end{description}}

\def\beq{\begin{equation}}
\def\eeq{\end{equation}}
\def\bea{\begin{eqnarray}}
\def\eea{\end{eqnarray}}
\def\beas{\begin{eqnarray*}}
\def\eeas{\end{eqnarray*}}

\newtheorem{theorem}{Theorem}[section]
\newtheorem{corollary}{Corollary}[section]

\newtheorem{example}{Example}[section]

\newcommand{\qm}[1]{``#1''}
\newcommand{\M}{\mathcal{M}}

\newcommand{\seq}[1]{\left<#1\right>}

\newcommand{\nms}{\normalsize}

\begin{document}

\title{\textbf{Single-tape and Multi-tape Turing machines\\ through the lens of the
 Grossone methodology}\thanks{
This   research   was   supported   by  the  project \qm{High
accuracy supercomputations  and  solving global optimization
problems using the information  approach}  of the Russian Federal
Program \qm{Scientists and Educators  in  Russia  of  Innovations},
grant number 14.B37.21.0878.}}

\author{Yaroslav D. Sergeyev$^{1-3,}$\footnote{{\tt yaro@si.deis.unical.it}
 Corresponding author}\,\,\,  and
 Alfredo Garro$^{1,}$\footnote{{\tt alfredo.garro@unical.it}}\\[-4pt]
 \\
      \nms $^1$Universit\`a della Calabria,  87030 Rende (CS), Italy\\  [-4pt]
        \nms $^2$N.I.~Lobatchevsky State University,   Nizhni Novgorod, Russia  \\  [-4pt]
  \nms  $^3$Istituto di Calcolo e Reti ad Alte Prestazioni, C.N.R., Rende (CS), Italy
}

\date{}

\maketitle

\begin{abstract}
The paper investigates how  the mathematical languages used to
describe and to observe automatic computations influence  the
accuracy of the obtained results. In particular, we focus our
attention on Single and Multi-tape Turing machines which are
described and observed through the lens of a new mathematical
language which is strongly based on three methodological ideas
borrowed from Physics and applied to Mathematics, namely: the
distinction between the object (we speak here about a mathematical
object) of an observation and the instrument used for this
observation; interrelations holding between the object and the tool
used for the observation; the accuracy of the observation determined
by the tool. Results of the observation executed by the traditional
and new languages are compared and discussed.

\end{abstract}


\section{Introduction}
\label{s0_Turing}

Since the beginning of the last century,    the fundamental nature
of the concept of \textit{automatic computations} attracted a great
attention of mathematicians and computer scientists (see
\cite{Church,Kleene,Kolmogorov,Kolmogorov_Uspensky,Markov,Mayberry,Post,Turing}).
The first studies had as their reference context the David Hilbert
programme, and as their reference language that introduced by Georg
Cantor \cite{Cantor}. These approaches lead to different
mathematical models of computing machines (see
\cite{Ausiello,Barry_Cooper,Davis}) that, surprisingly, were
discovered to be equivalent (e.g., anything computable in the
$\lambda$-calculus is computable by a Turing machine). Moreover,
these results, and expecially those obtained by Alonzo Church, Alan
Turing \cite{Church,Godel_1931,Turing} and Kurt G\"odel, gave
fundamental contributions to demonstrate that David Hilbert
programme, which was based on the idea that all of the Mathematics
could be precisely axiomatized, cannot be realized.

In spite of this fact, the idea of finding  an adequate set of
axioms for one or another field of  Mathematics continues to be
among the most attractive goals for contemporary mathematicians.
Usually, when it is necessary  to define a concept or an object,
logicians try to introduce a number of axioms describing the object
in the absolutely best way. However, it is not clear how to reach
this absoluteness; indeed, when we describe a mathematical object or
a concept we are limited by the expressive capacity of the language
we use to make this description. A richer language allows us to say
more about the object and a weaker language -- less. Thus, the
continuous development of the mathematical (and not only
mathematical) languages leads to a continuous necessity of a
transcription and specification of axiomatic systems. Second, there
is no guarantee that the chosen axiomatic system defines
`sufficiently well' the required concept and a continuous comparison
with practice is required in order to check the goodness of the
accepted set of axioms. However, there cannot be again any guarantee
that the new version will be the last and definitive one. Finally,
the third limitation already mentioned above has been discovered by
G\"odel in his two famous incompleteness theorems
(see~\cite{Godel_1931}).

Starting from these considerations, in this paper,  we study the
relativity of mathematical languages in situations where they are
used to observe and to describe automatic computations. We consider
the traditional computational paradigm  mainly following results of
Turing (see \cite{Turing}) whereas emerging computational paradigms
(see, e.g.,
~\cite{Adamatzky,Nielsen,Zilinskas_informatica,Walster_1}) are not
considered here. In particular, we focus our attention on different
kinds of Turing machines by enriching and extending the results
presented in \cite{Sergeyev_Garro}.

The point of view presented in this paper uses strongly  three
methodological ideas borrowed from Physics and applied to
Mathematics, namely: the distinction between the object (we speak
here about a mathematical object) of an observation and the
instrument used for this observation; interrelations holding between
the object and the tool used for this observation; the accuracy of
the observation determined by the tool.

The main attention is dedicated to numeral systems\footnote{We are
reminded that a \textit{numeral} is a symbol or group of symbols
that represents a \textit{number}. The difference between numerals
and numbers is the same as the difference between words and the
things they refer to. A \textit{number} is a concept that a
\textit{numeral} expresses. The same number can be represented by
different numerals. For example, the symbols `7', `seven', and `VII'
are different numerals, but they all represent the same number.}
that we use to write down numbers, functions, models, etc. and that
are among our tools of investigation of mathematical and physical
objects. It is shown that numeral systems strongly influence our
capabilities to describe both the mathematical and physical worlds.
A new numeral system introduced in
\cite{Sergeyev,informatica,Lagrange}) for performing computations
with infinite and infinitesimal quantities is used for the
observation of mathematical objects and studying Turing machines.
The new methodology is based on the principle `The part is less than
the whole' introduced by Ancient Greeks and observed in practice. It
is applied to all sets and processes (finite and infinite) and all
numbers (finite, infinite, and infinitesimal).

In order to see the place of the new approach in the  historical
panorama of ideas dealing with infinite and infinitesimal, see
\cite{Lolli,MM_bijection,Dif_Calculus,first,Sergeyev_Garro}. The new
methodology has been successfully applied for studying a number of
applications: percolation (see \cite{Iudin,DeBartolo}), Euclidean
and hyperbolic geometry (see \cite{Margenstern,Rosinger2}), fractals
(see \cite{chaos,Menger,Biology,DeBartolo}), numerical
differentiation and optimization (see
\cite{DeLeone,Korea,Num_dif,Zilinskas}), infinite series (see
\cite{Dif_Calculus,Riemann,Zhigljavsky}), the first Hilbert problem
(see \cite{first}), and cellular automata (see \cite{DAlotto}).

The rest of the paper is structured as follows. In
Section~\ref{s1_Turing}, Single and Multi-tape Turing machines are
introduced along with \qm{classical} results concerning their
computational power and related equivalences; in
Section~\ref{s2_Turing} a brief introduction to the new language and
methodology is given whereas their exploitation for analyzing and
observing the different types of Turing machines is discussed in
Section~\ref{s3_Turing}. It shows that the new approach allows us to
observe Turing machines with a higher accuracy giving so the
possibility to better characterize and distinguish machines which
are equivalent when observed within the classical framework.
Finally, Section~\ref{s4_Turing} concludes the paper.

\section{Single and Multi-tape Turing Machines}
\label{s1_Turing}

The Turing machine is one of the  simple abstract computational
devices that can be used to model computational processes and
investigate the limits of computability. In the following
Subsections~\ref{s11_Turing} and~\ref{s12_Turing}, Single and
Multi-tape Turing machines will be described along with important
classical results concerning their computational power and related
equivalences.

\subsection{Single Tape Turing Machines}
\label{s11_Turing}

A Turing Machine (see, e.g.,
\cite{Hopcroft_Ullman,Turing}) can be defined as a 7-tuple
 \beq
\M=\seq{Q, \Gamma, \bar{b}, \Sigma, q_0, F, \delta},
\label{Turing_7}
 \eeq
 where $Q$ is a finite and not empty set of states; $\Gamma$ is a finite set of symbols; $\bar{b}\in\Gamma$ is a symbol called blank; $\Sigma \subseteq \{\Gamma-{\bar{b}}\}$ is the set of input/output
symbols; $q_0\in Q$ is the initial state; $F\subseteq Q$ is the set of final states; $\delta: \{Q-F\}\times\Gamma\mapsto Q\times\Gamma\times\{R,L,N\}$ is a partial function called the transition function, where $L$ means left, $R$ means right, and $N$ means no move.

Specifically, the machine is supplied with: (i) a \textit{tape} running through it which is divided into cells each capable of
containing a symbol $\gamma\in\Gamma $, where $\Gamma$ is called the tape alphabet, and $\bar{b}\in\Gamma$ is the only symbol
allowed to occur on the tape infinitely often; (ii) a \textit{head} that can read and write symbols on the tape and move
the tape left and right one  and only one  cell at a time. The behavior of the machine  is specified by its \textit{transition
function} $\delta$ and consists of a sequence of  computational steps; in each step the machine reads the symbol under the head
and applies the \textit{transition function} that, given the current state of the machine and the symbol it is reading on the
tape, specifies (if it is defined for these inputs): (i) the symbol $\gamma\in\Gamma$ to write on the cell of the tape under
the head; (ii) the move of the tape ($L$ for one cell left, $R$ for one cell right, $N$ for no move); (iii) the next state $q\in
Q$ of the machine.

Starting from the definition of Turing Machine introduced above,
classical results (see, e.g., \cite{Ausiello}) aim at showing that
different machines in terms of provided tape and alphabet have the
same computational power, i.e., they are able to execute the same
computations. In particular, two main results are reported below in
an informal way.

Given a  Turing Machine $\M=\{Q, \Gamma, \bar{b}, \Sigma, q_0, F,
\delta\}$, which is supplied with an infinite tape, it is always
possible to define a Turing Machine $\M'=\{Q', \Gamma', \bar{b},
\Sigma', q_0', F', \delta'\}$ which is supplied with a semi-infinite
tape (e.g., a tape with a left boundary) and is equivalent to $\M$,
i.e., is able to execute all the computations of $\M$.

Given a  Turing Machine $\M=\{Q, \Gamma, \bar{b}, \Sigma, q_0, F,
\delta\}$, it is always possible to define a Turing Machine
$\M'=\{Q', \Gamma', \bar{b}, \Sigma', q_0', F', \delta'\}$ with
$\left|\Sigma'\right|=1$ and $\Gamma'=\Sigma'\cup\{\bar{b}\}$, which
is equivalent to $\M$, i.e., is able to execute all the computations
of $\M$.

It should be mentioned that these results, together with the usual
conclusion regarding the equivalences of Turing machines, can be
interpreted in the following, less obvious, way: they show that when
we observe Turing machines by exploiting the classical framework we
are not able to distinguish, from the computational point of view,
Turing machines which are provided with alphabets having different
number of symbols and/or different kind of tapes (infinite or
semi-infinite) (see \cite{Sergeyev_Garro} for a detailed
discussion).

\subsection{Multi-tape Turing Machines}
\label{s12_Turing}

Let us consider a variant of the Turing Machine defined
in~(\ref{Turing_7}) where a machine is equipped with multiple tapes
that can be simultaneously accessed and updated through multiple
heads (one per tape). These machines can be used for a more direct
and intuitive resolution of different kind of computational
problems. As an example, in checking if a string is palindrome it
can be useful to have two tapes on which represent the input string
so that the verification can be efficiently performed by reading a
tape from left to right and the other one from right to left.

Moving towards a more formal definition, a $k$-tapes, $k\geq2$,
Turing machine (see~\cite{Hopcroft_Ullman}) can be defined (cf.
(\ref{Turing_7})) as a 7-tuple
 \beq
  \M_{K}=\seq{Q, \Gamma, \bar{b}, \Sigma,
  q_0, F, \delta^{(k)}},
 \label{Turing_9}
 \eeq
 where $\Sigma=\bigcup^{k}_{i=1}\Sigma_{i}$ is given by the union of the symbols in the k input/output alphabets $\Sigma_{1},\dots,\Sigma_{k}$; $\Gamma=\Sigma\cup\{\bar{b}\}$ where $\bar{b}$ is a symbol called blank; $Q$ is a finite and not empty set of states; $q_0\in Q$ is the initial state; $F\subseteq Q$ is the set of final
states; $\delta^{(k)}: \{Q-F\}\times\Gamma_{1}\times\dots\times\Gamma_{k}\mapsto Q\times\Gamma_{1}\times\dots\times\Gamma_{k}\times\{R,L,N\}^{k}$ is a partial function called the transition function, where $\Gamma_{i}=\Sigma_{i}\cup\{\bar{b}\}, i=1,\dots,k$, $L$ means left, $R$ means right, and $N$ means no move.

This definition of $\delta^{(k)}$ means that the machine executes a transition starting from an internal state $q_{i}$ and with the $k$ heads (one for each tape) above the characters ${a_{i}}_{1},\dots,{a_{i}}_{k}$, i.e., if $\delta^{(k)}(q_{1},{a_{i}}_{1},\dots,{a_{i}}_{k})=(q_{j},{a_{j}}_{1},\dots,{a_{j}}_{k},{z_{j}}_{1},\dots,{z_{j}}_{k})$ the machine goes in the new state $q_{j}$, write on the k tapes the characters ${a_{j}}_{1},\dots,{a_{j}}_{k}$ respectively, and moves each of its k heads left, right or no move, as specified by the ${z_{j}}_{l}\in\{R,L,N\}, l=1,\dots,k$.

A machine can adopt for each tape a different alphabet, in any case,
for each tape, as for the Single-tape Turing machines, the minimum
portion containing characters distinct from $\bar{b}$ is usually
represented. In general, a typical configuration of a Multi-tape
machine consists of a read-only input tape, several read and write
work tapes, and a write-only output tape, with the input and output
tapes accessible only in one direction. In the case of a $k$-tapes
machine, the instant configuration of the machine, as for the Single-tape
case, must describe the internal state, the contents of the
tapes and the positions of the heads of the machine.

More formally, for a $k$-tapes Turing machine $\M_{K}=\seq{Q,
\Gamma, \bar{b}, \Sigma, q_0, F, \delta^{(k)}}$ with
$\Sigma=\bigcup^{k}_{i=1}\Sigma_{i}$ (see~\ref{Turing_9}) a
configuration of the machine is given by: \beq
  q\#\alpha_{1}\uparrow\beta_{1}\#\alpha_{2}\uparrow\beta_{2}\#\dots\#\alpha_{k}\uparrow\beta_{k},
 \label{Turing_10}
 \eeq
where $q\in Q$; $\alpha_{i}\in \Sigma_{i}\Gamma^{*}_{i}\cup\{\epsilon\}$ and $\beta_{i}\in \Gamma^{*}_{i}\Sigma_{i}\cup\{\bar{b}\}$.
A configuration is \textit{final} if $q\in F$.

The \textit{starting} configuration usually requires the input
string $x$ on a tape, e.g., the first tape so that $x\in
\Sigma_{1}^{*}$, and only $\bar{b}$ symbols on all the other tapes.
However, it can be useful to assume that, at the beginning of a
computation, these tapes have a starting symbol
$Z_{0}\notin\Gamma=\bigcup^{k}_{i=1}\Gamma_{i}$. Therefore, in the
initial configuration the head on the first tape will be on the
first character of the input string $x$, whereas the heads on the
other tapes will observe the symbol $Z_{0}$, more formally, by
re-placing $\Gamma_{i}=\Sigma_{i}\cup\{\bar{b}, Z_{0}\}$ in all the
previous definition, a configuration
$q\#\alpha_{1}\uparrow\beta_{1}\#\alpha_{2}\uparrow\beta_{2}\#\dots\#\alpha_{k}\uparrow\beta_{k}$
is an \textit{initial configuration} if $\alpha_{i}=\epsilon,
i=1,\dots,k, \beta_{1}\in \Sigma_{1}^{*},\beta_{i}=Z_{0},
i=2,\dots,k$ and $q=q_{0}$.

 The application of the transition
function $\delta^{(k)}$ at a machine configuration (c.f.
(\ref{Turing_10})) defines a \textit{computational step} of a
Multi-tape Turing Machine. The set of computational steps which
bring the machine from the initial configuration into a final
configuration defines the \textit{computation} executed by the
machine. As an example, the computation of a Multi-tape Turing
machine $\M_{K}$ which computes the function $f_{\M_{K}}(x)$ can be
represented as follows: \beq
  q_{0}\#\uparrow x\#\uparrow Z_{0}\#\dots\#\uparrow Z_{0}\stackrel{\rightarrow}{\M_{K}}q\#\uparrow x\#\uparrow f_{\M_{K}}(x)\#\uparrow \bar{b}\#\dots\#\uparrow\bar{b}
 \label{Turing_11}
 \eeq
where $q \in F$ and $\stackrel{\rightarrow}{\M_{K}}$ indicates the transition among machine configurations.

 It is worth noting that, although the $k$-tapes
Turing Machine can be used for a more direct resolution of different
kind of computational problems, in the classical framework it has
the same computational power of the Single-tape Turing machine. More
formally, given a Multi-tape Turing Machine it is always possible to
define a Single-tape Turing Machine which is able to fully simulate
its behavior and therefore to completely execute its computations.
In particular, the Single-tape Turing Machines adopted for the
simulation use a particular kind of the tape which is divided into
tracks (multi-track tape). In this way, if the tape has $m$ tracks,
the head is able to access (for reading and/or writing) all the $m$
characters on the tracks during a single operation. If for the $m$
tracks the alphabets $\Gamma_{1},\dots\,\Gamma_{m}$ are adopted
respectively, the machine alphabet $\Gamma$ is such that
$\left|\Gamma\right|=\left|\Gamma_{1} \times \dots \times
\Gamma_{m}\right|$ and can be defined by an injective function from
the set $\Gamma_{1} \times \dots \times \Gamma_{m}$ to the set
$\Gamma$; this function will associate the symbol $\bar{b}$ in
$\Gamma$ to the tuple $(\bar{b},\bar{b},\dots,\bar{b})$ in
$\Gamma_{1} \times \dots \times \Gamma_{m}$. In general, the
elements of $\Gamma$ which correspond to the elements in $\Gamma_{1}
\times \dots \times \Gamma_{m}$ can be indicated by $[
{a_{i}}_{1},{a_{i}}_{2},\dots,{a_{i}}_{m} ]$ where ${a_{i}}_{j} \in
\Gamma_{j}$.

By adopting this notation it is possible to demonstrate that given a
$k$-tapes Turing Machine $\M_{K}=\{ Q, \Gamma, \bar{b}, \Sigma, q_0,
F, \delta^{(k)}\}$ it is always possible to define a Single-tape
Turing Machine which is able to simulate $t$ computational steps of
$\M_{K}=$ in $O(t^{2})$ transitions by using an alphabet with
$O((2\left|\Gamma\right|)^{k})$ symbols (see \cite{Ausiello}).

The proof is based on the definition of a machine $\M'=\{ Q',
\Gamma', \bar{b}, \Sigma', q_0', F', \delta' \}$ with a Single-tape
divided into $2k$ tracks (see \cite{Ausiello}); $k$ tracks for
storing the characters in the $k$ tapes of $\M_{K}$ and $k$ tracks
for signing through the marker $\downarrow$ the positions of the $k$
heads on the $k$ tapes of $\M_{k}$. As an example, this kind of tape
can represent the content of each tapes of $\M_{k}$ and the position
of each machine heads in its even and odd tracks respectively. As
discussed above, for obtaining a Single-tape machine able to
represent these $2k$ tracks, it is sufficient to adopt an alphabet
with the required cardinality and define an injective function which
associates a 2k-ple characters of a cell of the multi-track tape to
a symbols in this alphabet.

The transition function $\delta^{(k)}$ of the $k$-tapes machine is
given by
$\delta^{(k)}(q_{1},{a_{i}}_{1},\dots,{a_{i}}_{k})=(q_{j},{a_{j}}_{1},\dots,{a_{j}}_{k},{z_{j}}_{1},\dots,{z_{j}}_{k})$,
with ${z_{j}}_{1}, \dots,{z_{j}}_{k} \in\{R,L,N\}$; as a consequence
the corresponding transition function $\delta'$ of the Single-tape
machine, for each transition specified by $\delta^{(k)}$ must
individuate the current state and the position of the marker for
each track and then write on the tracks the required symbols, move
the markers and go in another internal state. For each computational
step of $\M_{K}$, the machine $\M'$ must execute a sequence of steps
for covering the portion of tapes between the two most distant
markers. As in each computational step a marker can move at most of
one cell and then two markers can move away each other at most of
two cells, after $t$ steps of $\M_{K}$ the markers can be at most
$2t$ cells distant, thus if $\M_{K}$ executes $t$ steps, $\M'$
executes at most: $2\sum^{t}_{i=1}i = t^{2}+t =O(t^{2})$ steps.

Moving to the cost of the simulation in terms of the number of
required characters for the alphabet of the Single-tape machine, we
recall that $\left|\Gamma_{1}\right|=\left|\Sigma_{1}\right|+1$ and
that $\left|\Gamma_{i}\right|=\left|\Sigma_{i}\right|+2$ for $2\leq
i\leq k$. So by multiplying the cardinalities of these alphabets we
obtain that:
$\left|\Gamma'\right|=2^{k}(\left|\Sigma_{1}\right|+1)\prod^{k}_{i=2}(\left|\Sigma_{i}\right|+2)=O({(2{max}_{1\leq
i\leq k}\left|\Gamma_{i}\right|)}^{k})$.

\section{The Grossone Methodology}
\label{s2_Turing}

In this section, we give just a brief introduction to the
methodology of the new approach \cite{Sergeyev,informatica} dwelling
only on the issues directly related to the subject of the paper.
This methodology will be used in Section~\ref{s3_Turing} to study
Turing machines and to obtain some more accurate results with
respect to those obtainable by using the traditional framework
\cite{Church,Turing}.

In order to start, let us remind that numerous trials have been done
during the centuries to evolve existing numeral systems in such a
way that numerals representing infinite and infinitesimal numbers
could be included in them (see
\cite{Benci,Cantor,Conway,Leibniz,Levi-Civita,Newton,Robinson,Wallis}).
Since new numeral systems appear very rarely, in each concrete
historical period their significance  for Mathematics is very often
underestimated (especially by pure mathematicians). In order to
illustrate their importance, let us remind the Roman numeral system
that does not allow one to express zero and negative numbers. In
this system, the expression III-X is an indeterminate form. As a
result, before appearing the positional numeral system and inventing
zero mathematicians were not able to create theorems involving zero
and negative numbers and to execute computations with them.

There exist numeral systems that are even weaker than the Roman one. They seriously limit their users in executing computations. Let us recall a study published recently in \textit{Science} (see \cite{Gordon}). It describes a primitive tribe living in Amazonia (Pirah\~{a}). These people use a very simple numeral system for counting: one, two, many. For Pirah\~{a}, all quantities larger than two are just `many' and such operations as 2+2 and 2+1 give the same result, i.e., `many'. Using their weak numeral system Pirah\~{a} are not able to see, for instance, numbers 3, 4, 5, and 6, to execute arithmetical operations with them, and, in general, to say anything about these numbers because in their language there are neither words nor concepts for that.

In the context of the present paper, it is very important that the weakness of Pirah\~{a}'s numeral system leads them to such results as
  \beq
  \mbox{`many'}+ 1= \mbox{`many'},   \hspace{1cm}
\mbox{`many'} + 2 = \mbox{`many'},
   \label{piraha1}
        \eeq
which are very familiar to us  in the context of views on infinity used in the traditional calculus
  \beq
   \infty + 1= \infty,
\hspace{1cm}    \infty + 2 = \infty.
  \label{piraha2}
        \eeq
The arithmetic of Pirah\~{a} involving the numeral `many' has also a
clear similarity with the arithmetic proposed by Cantor for his
Alephs\footnote{This similarity becomes even more pronounced  if one
considers another Amazonian  tribe -- Munduruk\'u (see \cite{Pica})
-- who fail in exact arithmetic with numbers larger than 5 but are
able to compare and add large approximate numbers that are far
beyond their naming range. Particularly, they use the words `some,
not many' and `many, really many' to distinguish two types of large
numbers using the rules that are very similar to ones used by Cantor
to operate with $\aleph_0$ and $\aleph_1$, respectively.}:
  \beq
\aleph_0 + 1= \aleph_0,    \hspace{1cm} \aleph_0 + 2= \aleph_0,
\hspace{1cm}\aleph_1+ 1= \aleph_1,    \hspace{1cm}   \aleph_1 + 2 =
\aleph_1.
   \label{piraha3}
        \eeq

Thus, the modern mathematical numeral systems allow us to distinguish a larger quantity of finite numbers with respect to Pirah\~{a} but give results that are similar to those of Pirah\~{a} when we speak about infinite quantities. This observation leads us to the following idea: \textit{Probably our difficulties in working with infinity is not connected to the nature of infinity itself but is a result of inadequate numeral systems that we use to work with infinity, more precisely, to express infinite numbers.}

The approach developed  in \cite{Sergeyev,informatica,Lagrange}
proposes a numeral system that uses  the same numerals for several
different purposes for dealing with infinities and infinitesimals:
in Analysis for working with functions that can assume different
infinite, finite, and infinitesimal values (functions can also have
derivatives assuming different infinite or infinitesimal values);
for measuring infinite sets; for indicating positions of elements in
ordered infinite sequences; in probability theory, etc. (see
\cite{DeLeone,DAlotto,Iudin,Margenstern,Rosinger2,chaos,Menger,Korea,Dif_Calculus,first,Num_dif,Riemann,Biology,DeBartolo,Zhigljavsky,Zilinskas}).
It is important to emphasize that the new numeral system   avoids
situations of the type (\ref{piraha1})--(\ref{piraha3}) providing
results ensuring that  if $a$ is a numeral written in this system
then for any $a$ (i.e., $a$ can be finite, infinite, or
infinitesimal) it follows $a+1>a$.

The new numeral system works as follows. A new infinite unit of
measure expressed by the numeral \ding{172} called \textit{grossone}
is introduced as the number of elements of the set, $\mathbb{N}$, of
natural numbers. Concurrently with the introduction of grossone in
the mathematical language all other symbols (like $\infty$, Cantor's
$\omega$, $\aleph_0, \aleph_1, ...$,  etc.) traditionally used to
deal  with infinities and infinitesimals are excluded from the
language because grossone and other numbers constructed with its
help not only can be used instead of all of them but  can be used
with a higher accuracy\footnote{Analogously, when the switch from
Roman numerals to the Arabic ones has been done, numerals X, V, I, etc.
have been excluded from records using Arabic numerals.}.
Grossone is introduced by describing its properties postulated by
the Infinite Unit Axiom (see \cite{informatica,Lagrange}) added to
axioms for real numbers (similarly, in order to pass from the set,
$\mathbb{N}$, of natural numbers to the set, $\mathbb{Z}$, of
integers a new element -- zero expressed by the numeral 0 -- is
introduced by describing its properties).

The new numeral \ding{172} allows us to construct different numerals
expressing different infinite and infinitesimal numbers and to
execute computations with them. Let us give some examples.  For
instance, in Analysis, indeterminate forms are not present and, for
example, the following relations hold for $\mbox{\ding{172}}$ and
$\mbox{\ding{172}}^{-1}$ (that is infinitesimal), as for any other
(finite, infinite, or infinitesimal) number expressible in the new
numeral system
  \beq
  0 \cdot \mbox{\ding{172}} =
\mbox{\ding{172}} \cdot 0 = 0, \hspace{3mm} \mbox{\ding{172}}-\mbox{\ding{172}}= 0,\hspace{3mm} \frac{\mbox{\ding{172}}}{\mbox{\ding{172}}}=1, \hspace{3mm} \mbox{\ding{172}}^0=1, \hspace{3mm} 1^{\mbox{\tiny{\ding{172}}}}=1, \hspace{3mm} 0^{\mbox{\tiny{\ding{172}}}}=0,
  \label{3.2.1}
        \eeq
\beq
  0 \cdot \mbox{\ding{172}}^{-1} =
\mbox{\ding{172}}^{-1} \cdot 0 = 0, \hspace{5mm} \mbox{\ding{172}}^{-1} > 0, \hspace{5mm} \mbox{\ding{172}}^{-2} > 0, \hspace{5mm} \mbox{\ding{172}}^{-1}-\mbox{\ding{172}}^{-1}= 0,
\label{3.2.1.1}
  \eeq
\beq
 \frac{\mbox{\ding{172}}^{-1}}{\mbox{\ding{172}}^{-1}}=1,
\hspace{3mm}
\frac{\mbox{\ding{172}}^{-2}}{\mbox{\ding{172}}^{-2}}=1,
\hspace{3mm} (\mbox{\ding{172}}^{-1})^0=1, \hspace{5mm} \mbox{\ding{172}} \cdot \mbox{\ding{172}}^{-1} =1, \hspace{5mm} \mbox{\ding{172}} \cdot \mbox{\ding{172}}^{-2} =\mbox{\ding{172}}^{-1}.
      \label{3.2.1.2}
  \eeq

The new approach gives the possibility to develop a new Analysis (see  \cite{Dif_Calculus}) where  functions assuming not only finite values but also infinite and infinitesimal ones can be studied. For all of them it becomes possible to introduce a new notion of continuity that is closer to our modern physical knowledge.
Functions assuming finite and infinite values can be differentiated and integrated.

By using the new numeral system it becomes possible to measure certain infinite sets and to see, e.g., that the sets
of even and odd numbers have \ding{172}$/2$ elements each. The set, $\mathbb{Z}$, of integers has $2$\ding{172}$+1$ elements (\ding{172} positive elements, \ding{172} negative elements, and zero). Within the countable sets and sets having cardinality of the continuum (see \cite{Lolli,first,Lagrange}) it becomes possible to distinguish infinite sets having different number of elements expressible in the numeral system using grossone and to see that, for instance,
\[
  \frac{\mbox{\ding{172}}}{2} < \mbox{\ding{172}}-1 < \mbox{\ding{172}} < \mbox{\ding{172}}+1 < 2\mbox{\ding{172}}+1 <
  2\mbox{\ding{172}}^2-1 < 2\mbox{\ding{172}}^2    <
  2\mbox{\ding{172}}^2+1  < \]
\beq
2\mbox{\ding{172}}^2+2  <  2^{\mbox{\ding{172}}}-1 < 2^{\mbox{\ding{172}}} < 2^{\mbox{\ding{172}}}+1 < 10^{\mbox{\ding{172}}} <
   \mbox{\ding{172}}^{\mbox{\ding{172}}}-1 <
   \mbox{\ding{172}}^{\mbox{\ding{172}}} <
   \mbox{\ding{172}}^{\mbox{\ding{172}}}+1.
 \label{dis}
        \eeq

Another key notion for our study of Turing machines is that of infinite sequence. Thus, before considering the notion of the Turing machine from the point of view of the new methodology, let us explain how the notion of the infinite sequence can be viewed from the new positions.

Traditionally, an \textit{infinite sequence} $\{a_n\}, a_n \in A,$
$n \in \mathbb{N},$ is defined as a function having the set of
natural numbers, $\mathbb{N}$, as the domain  and a set $A$ as the
codomain. A \textit{subsequence} $\{b_n\}$ is defined as a sequence
$\{a_n\}$ from which some of its elements have been removed. In
spite of the fact that the removal of the elements from $\{a_n\}$
can be directly observed, the traditional approach does not allow
one to register, in the case where the obtained subsequence
$\{b_n\}$ is infinite,   the fact that $\{b_n\}$ has less elements
than the original infinite sequence $\{a_n\}$.

Let us study what happens when the new  approach is used. From the
point of view of the new methodology, an infinite sequence can be
considered in a dual way: either as an object of a mathematical
study or as a mathematical instrument developed by human beings to
observe other objects and processes.  First, let us consider it as a
mathematical object and show that the definition of infinite
sequences should be done more precise within the new methodology. In
the finite case, a sequence $a_1, a_2, \ldots , a_n$ has $n$
elements and we extend this definition directly to the infinite case
saying that an infinite sequence $a_1, a_2, \ldots , a_n$ has $n$
elements where $n$ is expressed by an infinite numeral such that the
operations with it satisfy  the methodological Postulate~3. Then the
following result (see \cite{Sergeyev,informatica}) holds. We
reproduce here its proof for the sake of completeness.
\begin{theorem}
\label{t2} The number of  elements of any infinite sequence is less
or equal to~\ding{172}.
\end{theorem}

\textit{Proof.}  The new numeral system allows us to express the number of elements of the set $\mathbb{N}$  as \ding{172}. Thus, due to the sequence definition given above, any sequence having $\mathbb{N}$ as the domain  has \ding{172} elements.

The notion of subsequence is introduced as a sequence from which some of its elements have been removed. This means that the resulting subsequence will have less elements than the original sequence. Thus, we obtain infinite sequences having the number of members less than grossone.  \hfill $\Box$

It becomes appropriate now to define the \textit{complete sequence} as an infinite sequence  containing \ding{172} elements.
For example, the sequence   of natural numbers  is complete, the sequences of even  and odd natural numbers  are not complete
because  they have $\frac{\mbox{\ding{172}}}{2}$ elements each (see \cite{Sergeyev,informatica}). Thus, the new approach imposes
a more precise description of infinite sequences than the traditional one: to define a sequence $\{a_n\}$ in the new language, it is not sufficient just to give a formula for~$a_n$, we should determine (as it happens for sequences having a finite number of elements) its number of elements and/or the first and the last elements of the sequence. If the number of the first element is equal to one, we can use the record $\{a_n: k \}$ where $a_n$ is, as usual, the general element of the sequence and $k$ is the number (that can be finite or infinite) of members of the sequence; the following example clarifies these concepts.

 \begin{example} \label{e0_Turing}
Let us consider the following three sequences:
\beq
 \{a_n:\mbox{\ding{172}}\} =  \{ 4,\hspace{3mm} 8,\hspace{3mm} \ldots \hspace{3mm} 4(\mbox{\ding{172}}-1),\hspace{3mm}
4\mbox{\ding{172}} \};
\eeq
\beq \{b_n:\frac{\mbox{\ding{172}}}{2}-1\}  = \{  4,\hspace{3mm}8,\hspace{3mm} \ldots
\hspace{3mm} 4(\frac{\mbox{\ding{172}}}{2}-2),\hspace{3mm}
4(\frac{\mbox{\ding{172}}}{2}-1) \};
 \eeq
\beq
 \{c_n:\frac{\mbox{2\ding{172}}}{3}\}  = \{ 4,\hspace{3mm} 8,\hspace{3mm} \ldots \hspace{3mm}
4(\frac{\mbox{2\ding{172}}}{3}-1),\hspace{3mm} 4
\frac{2\mbox{\ding{172}}}{3}  \}.
 \eeq
The three sequences have $a_n=b_n=c_n=4n$ but they  are different because they have different number of members. Sequence
$\{a_n\}$ has \ding{172} elements and, therefore, is complete,
 $\{b_n\}$
has $\frac{\mbox{\ding{172}}}{2}-1$, and
   $\{c_n\}$ has
$2\frac{\mbox{\ding{172}}}{3}$ elements.
\hfill$\Box$
\end{example}

Let us consider now infinite sequences as one of the instruments used by mathematicians to study the world around us and other
mathematical objects and processes. The first immediate consequence of Theorem~\ref{t2} is that any \textit{sequential} process can have at maximum \ding{172} elements. This means that a process of sequential observations of any object cannot
contain more than \ding{172} steps\footnote{It is worthy to notice a deep relation of this observation to the Axiom of Choice. Since Theorem~\ref{t2} states that any sequence can have at maximum \ding{172} elements, so this fact holds for the process of a
sequential choice, as well. As a consequence, it is not possible to choose sequentially more than \ding{172} elements from a set. This observation also emphasizes the fact that the parallel computational paradigm is significantly different with respect to the sequential one because $p$ parallel processes can choose $p\cdot \mbox{\ding{172}}$ elements from a set.}.
We are not able to execute any infinite process physically but we assume the existence of such a process; moreover, only a finite number of observations of elements of the considered infinite sequence can be executed by a human who is limited by the numeral system used for the observation. Indeed, we can observe only those members of a sequence for which there exist the corresponding numerals in the chosen numeral system; to better clarify this point the following example is discussed.

\begin{example} \label{e3_Turing}  Let us consider the numeral system, $\mathcal{P}$, of Pirah\~{a} able to express only numbers 1 and 2. If we add to $\mathcal{P}$ the new numeral \ding{172}, we obtain a new numeral system (we call it $\widehat{\mathcal{P}}$). Let us consider now a sequence of natural numbers $\{n: \mbox{\ding{172}}
\}$. It goes from 1 to \ding{172} (note that both numbers, 1 and \ding{172}, can be expressed by numerals from
$\widehat{\mathcal{P}}$). However, the numeral system $\widehat{\mathcal{P}}$ is very weak and it allows us    to
observe only ten numbers from the sequence $\{n:
\mbox{\ding{172}}\}$ represented by the following numerals
 \beq \underbrace{1,2}_{finite},
\hspace{5mm} \ldots \hspace{5mm}
\underbrace{\frac{\mbox{\small{\ding{172}}}}{2}-2,
\frac{\mbox{\small{\ding{172}}}}{2}-1,
\frac{\mbox{\small{\ding{172}}}}{2},
\frac{\mbox{\small{\ding{172}}}}{2}+1,
\frac{\mbox{\small{\ding{172}}}}{2}+2}_{infinite}, \hspace{5mm}
\ldots \hspace{5mm} \underbrace{\mbox{\ding{172}}-2,
\mbox{\ding{172}}-1, \mbox{\ding{172}}}_{infinite}.
 \label{grossone_10}
       \eeq
The first two numerals in (\ref{grossone_10})  represent finite
numbers, the remaining eight numerals express infinite numbers, and
dots represent members of the sequence of natural numbers that are
not expressible in~$\widehat{\mathcal{P}}$ and, therefore, cannot be
observed if one uses only this numeral system for this purpose.
\hfill$\Box$
\end{example}

In the light of the limitations  concerning the process of
sequential observations, the researcher can choose how to organize
the required sequence of observations and which numeral system to
use for it, defining so which elements of the object he/she can
observe. This situation is exactly the same as in natural sciences:
before starting to study a physical object, a scientist chooses an
instrument and its accuracy for the study.

\begin{example}
\label{e4_Turing} Let us consider the set A=$\{1,2,3,\ldots,2$\ding{172}-1,2\ding{172}$\}$ as an object of our observation. Suppose that we want to organize the process of the sequential counting of its elements. Then, due to
Theorem~\ref{t2}, starting from the number 1 this process can arrive at maximum to \ding{172}. If we consider the complete
counting sequence $\{n: \mbox{\ding{172}}\}$, then we obtain
 \beq
 \begin{array}{l}
              1,2,\hspace{1mm}3,\hspace{1mm}4, \hspace{2mm} \ldots \hspace{2mm}
              \mbox{\ding{172}\tiny$^-$}2,\mbox{\ding{172}\tiny$^-$}1,
              \mbox{\ding{172}}, \mbox{\ding{172}\tiny$^+$}1,
              \mbox{\ding{172}\tiny$^+$}2, \mbox{\ding{172}\tiny$^+$}3,\ldots,\mbox{2\ding{172}\tiny$^-$}1,\mbox{2\ding{172}}\\
              \hspace{1mm}\raisebox{3.5ex}{\scalebox{1.0}[1.0]{\rotatebox{180}{$\curvearrowleft$}}}
              \raisebox{3.5ex}{\scalebox{1.0}[1.0]{\rotatebox{180}{$\curvearrowleft$}}}
              \raisebox{3.5ex}{\scalebox{1.1}[1.0]{\rotatebox{180}{$\curvearrowleft$}}}
              \raisebox{3.5ex}{\scalebox{1.1}[1.0]{\rotatebox{180}{$\curvearrowleft$}}}
              \hspace{4.5mm}
              \raisebox{3.5ex}{\scalebox{1.5}[1.0]{\rotatebox{180}{$\curvearrowleft$}}}
              \hspace{1.3mm}
              \raisebox{3.5ex}{\scalebox{1.5}[1.0]{\rotatebox{180}{$\curvearrowleft$}}}
              \hspace{2mm}
              \raisebox{3.5ex}{\scalebox{1.5}[1.0]{\rotatebox{180}{$\curvearrowleft$}}}
              \hspace{0.8mm}
        \vspace*{-4.5mm}\\
              \raisebox{5.5ex}{\scalebox{1}{\rotatebox{0}{$\underbrace{ \hspace{40mm} }_{ \mbox{\ding{172} \small \,steps} } $}}}
 \end{array}
 \label{grossone_11}
       \eeq

Analogously, if we start  the process of the sequential counting from 5, the process   arrives at maximum to $\mbox{\ding{172}}+4$:
\beq
\begin{array}{l}
              1,2,3,4,5 \hspace{1mm} \ldots \hspace{2mm}
              \mbox{\ding{172}\tiny$^-$}1,
              \mbox{\ding{172}}, \mbox{\ding{172}\tiny$^+$}1,
              \mbox{\ding{172}\tiny$^+$}2, \mbox{\ding{172}\tiny$^+$}3, \mbox{\ding{172}\tiny$^+$}4,\mbox{\ding{172}\tiny$^+$}5, \ldots,\mbox{2\ding{172}\tiny$^-$}1,\mbox{2\ding{172}} \\
              \hspace{13.5mm}
              \raisebox{3.5ex}{\scalebox{1.1}[1.0]{\rotatebox{180}{$\curvearrowleft$}}}
              \hspace{1.5mm}
              \raisebox{3.5ex}{\scalebox{1.5}[1.0]{\rotatebox{180}{$\curvearrowleft$}}}
              \hspace{1.0mm}
              \raisebox{3.5ex}{\scalebox{1.5}[1.0]{\rotatebox{180}{$\curvearrowleft$}}}
              \hspace{0.5mm}
              \raisebox{3.5ex}{\scalebox{1.5}[1.0]{\rotatebox{180}{$\curvearrowleft$}}}
              \hspace{0.1mm}
              \raisebox{3.5ex}{\scalebox{2.1}[1.0]{\rotatebox{180}{$\curvearrowleft$}}}
              \hspace{0.2mm}
              \raisebox{3.5ex}{\scalebox{2.1}[1.0]{\rotatebox{180}{$\curvearrowleft$}}}
              \hspace{0.2mm}
              \raisebox{3.5ex}{\scalebox{2.1}[1.0]{\rotatebox{180}{$\curvearrowleft$}}}
              \vspace*{-4.5mm}\\
              \hspace{13mm}\raisebox{5.5ex}{\scalebox{1}{\rotatebox{0}{$\underbrace{ \hspace{49mm} }_{ \mbox{\ding{172} \small \,steps} } $}}}
 \end{array}
 \eeq
The corresponding complete sequence  used in this case is $\{n+4: \mbox{\ding{172}}\}$. We can also change the length of the step in the counting sequence and consider, for instance, the complete sequence $\{2n-1: \mbox{\ding{172}}\}$:
\beq
 \begin{array}{l}
              1,2,3,4,\hspace{1mm}  \ldots \hspace{1mm}
 \mbox{\ding{172}\tiny$^-$}1,
\mbox{\ding{172}},  \mbox{\ding{172}\tiny$^+$}1,
\mbox{\ding{172}\tiny$^+$}2,
 \hspace{1mm} \ldots \hspace{1mm} 2\mbox{\ding{172}\tiny$^-$}3, 2 \mbox{\ding{172}\tiny$^-$}2, 2\mbox{\ding{172}\tiny$^-$}1, 2 \mbox{\ding{172}}\\
              \hspace{1mm}
              \raisebox{3.5ex}{\scalebox{2}[1.1]{\rotatebox{180}{$\curvearrowleft$}}}
              \hspace{-0.5mm}
              \raisebox{3.5ex}{\scalebox{2}[1.1]{\rotatebox{180}{$\curvearrowleft$}}}
               \hspace{2.3mm}
              \raisebox{3.5ex}{\scalebox{2}[1.1]{\rotatebox{180}{$\curvearrowleft$}}}
              \raisebox{3.5ex}{\scalebox{2.8}[1.1]{\rotatebox{180}{$\curvearrowleft$}}}
              \hspace{0.5mm}
              \raisebox{3.5ex}{\scalebox{4}[1.1]{\rotatebox{180}{$\curvearrowleft$}}}
              \hspace{1.0mm}
              \raisebox{3.5ex}{\scalebox{2}[1.1]{\rotatebox{180}{$\curvearrowleft$}}}
              \hspace{0.1mm}
              \raisebox{3.5ex}{\scalebox{5.1}[1.1]{\rotatebox{180}{$\curvearrowleft$}}}
                     \vspace*{-4.5mm}\\
              \raisebox{5.5ex}{\scalebox{1}{\rotatebox{0}{$\underbrace{ \hspace{75mm} }_{ \mbox{\ding{172} \small \,steps} } $}}}
 \end{array}
 \eeq
If we use again the numeral system $\widehat{\mathcal{P}}$, then among finite numbers it allows us to see only number 1 because already the next number in the sequence, 3, is not expressible in $\widehat{\mathcal{P}}$. The last element of
the sequence is $2\mbox{\ding{172}}-1$ and $\widehat{\mathcal{P}}$ allows us to observe it.  \hfill
 $\Box$
  \end{example}

The introduced definition of the sequence allows us to work not only
with the first but with any element of any sequence if the element
of our interest is expressible in the chosen numeral system
independently whether the sequence under our study has a finite or
an infinite number of elements. Let us use this new definition for
studying infinite sets of numerals, in particular, for calculating
the number of points at the interval $[0,1)$ (see
\cite{Sergeyev,informatica}). To do this we need a definition of the
term `point'\index{point} and mathematical tools to indicate a
point.   If we accept (as is usually done in modern Mathematics)
that a \textit{point} $A$ belonging to the interval $[0,1)$ is
determined by a numeral $x$, $x \in \mathbb{S},$ called
\textit{coordinate of the point A} where $\mathbb{S}$ is a set of
numerals, then we can indicate the point $A$ by its coordinate  $x$
and we are able to execute the required calculations.

It is worthwhile to  emphasize  that giving this definition we have
not used the usual formulation ``\textit{$x$ belongs to the set,
$\mathbb{R}$, of real numbers}''. This has been done because we can
express coordinates only by numerals and different choices of
numeral systems lead to different sets of numerals and, as a result,
to different sets of numbers observable through the chosen numerals.
In fact, we can express coordinates only after we have fixed a
numeral system (our instrument of the observation) and this choice
defines which points we can observe, namely, points having
coordinates expressible by the chosen numerals. This situation is
typical for natural sciences where it is well known that instruments
influence the results of observations. Remind the work with a
microscope: we decide the level of the precision we need and obtain
a result which is dependent on the chosen level of accuracy. If we
need a more precise or a more rough answer, we change the lens of
our microscope.

We should decide now which numerals we shall use to express coordinates of the points. After this choice we can calculate the
number of numerals expressible in the chosen numeral system and, as a result, we obtain the number of points at the interval $[0,1)$. Different variants (see \cite{Sergeyev,informatica}) can be chosen depending on the precision level we want to obtain.  For instance,   we can choose a positional numeral system with a finite radix $b$ that allows us to work with numerals
 \beq
(0.a_{1} a_{2}  \ldots a_{(\mbox{\tiny\ding{172}}-1)}
a_{\mbox{\tiny\ding{172}}})_b, \hspace{5mm}  a_{i} \in \{ 0, 1,
\ldots b-2, b-1 \}, \hspace{3mm}  1 \le i \le \mbox{\ding{172}}.
 \label{grossone_12}
       \eeq
Then, the number of numerals (\ref{grossone_12}) gives us the number of points within the interval $[0,1)$ that can be expressed by these
numerals. Note that  a number using  the positional numeral system (\ref{grossone_12}) cannot have more than grossone digits
(contrarily to sets discussed in Example~\ref{e4_Turing}) because a numeral having $g>\mbox{\ding{172}}$ digits would not be
observable in a sequence. In this case ($g>\mbox{\ding{172}}$) such a record  becomes useless in sequential computations because it does not allow one
to identify  numbers entirely since $g-\mbox{\ding{172}}$ numerals remain non observed.

\begin{theorem}
\label{t1_Turing} If  coordinates of   points $x \in [0,1)$ are expressed by numerals (\ref{grossone_12}), then the number of the points $x$ over $[0,1)$  is equal to $b^{\mbox{\tiny\ding{172}}}$.
\end{theorem}

\textit{Proof.} In the numerals (\ref{grossone_12})   there is a sequence  of digits, $a_{1} a_{2} \ldots
a_{(\mbox{\tiny\ding{172}}-1)} a_{\mbox{\tiny\ding{172}}}$, used to express   the fractional part of the number. Due to the
definition of the sequence and Theorem~\ref{t2}, any infinite sequence can have at maximum \ding{172} elements. As a result,
there is \ding{172} positions on the right of the dot that can be filled in by one of the $b$ digits from the alphabet\index{alphabet} $\{ 0, 1, \ldots , b-1 \}$ that leads to $b^{\mbox{\tiny\ding{172}}}$ possible combinations. Hence, the positional numeral system using the numerals of the form (\ref{grossone_12}) can express $b^{\mbox{\tiny\ding{172}}}$ numbers. \hfill
 $\Box$
\begin{corollary}
\label{c1_Turing}

 The  number of   numerals
  \beq
(a_{1} a_{2} a_{3}  \ldots
a_{\mbox{\tiny\ding{172}}-2}a_{\mbox{\tiny\ding{172}}-1}
a_{\mbox{\tiny\ding{172}}})_b, \hspace{5mm} a_i \in \{ 0, 1,
\ldots b-2, b-1 \}, \hspace{5mm}  1 \le i \le \mbox{\ding{172}},
 \label{Turing_4}
       \eeq
expressing integers in the positional system with a finite radix $b$ in the alphabet $\{ 0, 1, \ldots b-2, b-1 \}$ is equal to $b^{\mbox{\tiny\ding{172}}}$.
\end{corollary}

\textit{Proof.}  The  proof is a straightforward consequence of Theorem~\ref{t1_Turing} and is so omitted. \hfill $\Box$

\begin{corollary}
\label{c2_Turing}

If  coordinates of   points $x \in (0,1)$ are expressed by numerals (\ref{grossone_12}), then the number of the
points $x$ over $(0,1)$  is equal to $b^{\mbox{\tiny\ding{172}}}-1$.
\end{corollary}

\textit{Proof.}  The  proof follows immediately from Theorem~\ref{t1_Turing}. \hfill
 $\Box$

Note that Corollary \ref{c2_Turing} shows that it becomes possible now to observe and to register the difference of the
number of elements of two infinite sets (the interval $[0,1)$ and the interval $(0,1)$, respectively) even when only one element
(the point 0, expressed by the numeral $0.00\ldots0$ with \ding{172} zero digits after the decimal point) has been excluded from the first set in order to obtain the second one.

\section{The Turing Machines observed through the lens of the Grossone Methodology}
\label{s3_Turing}

In this Section  the different types of Turing machines introduced
in Section~\ref{s1_Turing} are analyzed and observed by using as
instruments of the observation the Grossone language and methodology
presented in Section~\ref{s2_Turing}. In particular, new results for
Multi-tape Turing machines are presented and discussed.

Before starting the discussion,  it is useful to recall the main
results from the previous Section: (i) any infinite sequence can
have maximum \ding{172} elements; (ii) the elements which we are
able to observe in this sequence depend on the adopted numeral
system.

Then, in order to  be able to read and to understand the output of a
Turing machine, writing its output on the tape using an alphabet
$\Sigma$ containing $b$ symbols $\{ 0, 1, \ldots b-2, b-1 \}$ where
$b$ is a finite number, the researcher (the user) should know a
positional numeral system $\mathcal{U}$ with an alphabet $\{ 0, 1,
\ldots u-2, u-1 \}$ where $u \ge b$, otherwise the output cannot be
decoded by the user. Moreover, the researcher must be able to
observe a number of symbols at least equal to the maximal length of
the output sequence that can be computed by machine, otherwise the
user is not able to interpret the obtained result (see
\cite{Sergeyev_Garro} for a detailed discussion).

In this Section,  a first set of results aims to specify, with
higher accuracy with respect to that provided by the mathematical
language developed by Cantor and adopted by Turing, how and when the
computations performed by a Multi-tape Turing machine can be
observed in a sequence. Moreover, it is shown that the Grossone
language and methodology will allow us to perform a more accurate
investigation of situations interpreted traditionally like
equivalences among different Multi-tape machines and among Multi and
Single-tape machines.

\subsection{Observing computations performed by a Multi-tape Turing machine}
\label{s321_Turing}

Before starting to analyze the computations performed by a $k$-tapes
Turing machine (with $k\geq2$) $\M_{K}=\seq{Q, \Gamma, \bar{b},
\Sigma, q_0, F, \delta^{(k)}}$ (see (\ref{Turing_7}), Section
\ref{s12_Turing}), it is worth to make some considerations about the
process of observation itself in the light of the Grossone
methodology. As discussed above, if we want to observe the process
of computation performed by a Turing machine while it executes an
algorithm, then we have to execute observations of the machine in a
sequence of moments. In fact, it is not possible to organize a
continuous observation of the machine. Any instrument used for an
observation has its accuracy and there always be a minimal period of
time related to this instrument allowing one to distinguish two
different moments of time and, as a consequence, to observe (and to
register) the states of the object in these two moments. In the
period of time passing between these two moments the object remains
unobservable.

Since our observations are made in a sequence, the process of
observations can have at maximum $\mbox{\ding{172}}$ elements. This
means that inside a computational process it is possible to fix more
than grossone steps (defined in a way) but it is not possible to
count them one by one in a sequence containing more than grossone
elements. For instance, in a time interval $[0,1)$, up to
$b^{\mbox{\ding{172}}}$ numerals of the type (\ref{grossone_12}) can
be used to identify moments of time but not more than grossone of
them can be observed in a sequence. Moreover, it is important to
stress that any process itself, considered independently on the
researcher, is not subdivided in iterations, intermediate results,
moments of observations, etc. The structure of the language we use
to describe the process imposes what we can say about the process
(see \cite{Sergeyev_Garro} for a detailed discussion).

On the basis of the considerations made above, we should choose the
accuracy (granularity) of the process of the observation of a Turing
machine; for instance we can choose a single operation of the
machine such as reading a symbol from the tape, or moving the tape,
etc. However, in order to be close as much as possible to the
traditional results, we consider an application of the transition
function of the machine as our observation granularity (see Section
\ref{s1_Turing}).

 Moreover, concerning the output of the machine,
we consider the symbols written on all the k tapes of the machine by
using, on each tape $i$, with $1\leq i \leq k$, the alphabet
$\Sigma_i$ of the tape, containing $b_i$ symbols, plus the blank
symbol ($\bar{b}$). Due to the definition of complete sequence (see
Section~\ref{s2_Turing}) on each tape at least \ding{172} symbols
can be produced and observed. This means that on a tape $i$, after
the last symbols belonging to the tape alphabet $\Sigma_i$, if the
sequence is not complete (i.e., if it has less than \ding{172}
symbols) we can consider a number of blank symbols ($\bar{b}$)
necessary to complete the sequence. We say that we are considering a
\textit{complete output} of a $k$-tapes Turing machine when on each
tape of the machine we consider a complete sequence of symbols
belonging to $\Sigma_{i}\cup\{\bar{b}\}$.

\begin{theorem}
\label{tkt1_Turing} Let $\M_{K}=\seq{Q, \Gamma, \bar{b}, \Sigma,
q_0, F, \delta^{(k)}}$ be a $k$-tapes, $k\geq2$, Turing machine.
Then, a complete output of the machine will results in $k$\ding{172}
symbols.
\end{theorem}

\textit{Proof.} Due to the definition  of the complete sequence, on
each tape at maximum \ding{172} symbols can be produced and observed
and thus by considering a complete sequence on each of the k tapes
of the machine the complete output of the machine will result in
$k$\ding{172} symbols. \hfill $\Box$

Having proved  that a complete output that can be produced by a
$k$-tapes Turing machine results in $k$\ding{172} symbols, it is
interesting to investigate what part of the complete output produced
by the machine can be observed in a sequence taking into account
that it is not possible to observe in a sequence more than
\ding{172} symbols (see Section~\ref{s2_Turing}). As examples, we
can decide to make in a sequence one of the following observations:
(i) \ding{172} symbols on one among the $k$-tapes of the machine,
(ii) $\frac{\mbox{\ding{172}}}{k}$ symbols on each of the $k$-tapes
of the machine; (iii) $\frac{\mbox{\ding{172}}}{2}$  symbols on $2$
among the $k$-tapes of the machine, an so on.

\begin{theorem}
\label{tkt12_Turing} Let $\M_{K}=\seq{Q, \Gamma, \bar{b}, \Sigma,
q_0, F, \delta^{(k)}}$ be a $k$-tapes, $k\geq2$, Turing machine. Let
$M$ be the number of all possible complete outputs that can be
produced by $\M_{K}$. Then it follows $M=\prod^{k}_{i=1}
{(b_i+1)}^{\mbox{\tiny\ding{172}}}$.
\end{theorem}

\textit{Proof.} Due to the definition  of the complete sequence, on
each tape $i$,  with $1\leq i \leq k$, at maximum \ding{172} symbols
can be produced and observed by using the $b_i$ symbols of the
alphabet $\Sigma_i$ of the tape plus the blank symbol ($\bar{b}$);
as a consequence, the number of all the possible complete sequences
that can be produced and observed on a tape $i$ is
${(b_i+1)}^{\mbox{\tiny\ding{172}}}$. A complete output of the
machine is obtained by considering a complete sequence on each of
the the $k$-tapes of the machine, thus by considering all the
possible complete sequences that can be produced and observed on
each of the k tapes of the machine, the number $M$ of all the
possible complete outputs will results in  $\prod^{k}_{i=1}
{(b_i+1)}^{\mbox{\tiny\ding{172}}}$. \hfill $\Box$

As the  number $M=\prod^{k}_{i=1}
{(b_i+1)}^{\mbox{\tiny\ding{172}}}$ of complete outputs that can be
produced by $\M_{K}$ is larger than grossone, then there can be
different sequential enumerating processes that enumerate complete
outputs in different ways, in any case, each of these enumerating
sequential processes cannot contain more than grossone members (see
Section~\ref{s2_Turing}).

\subsection{Equivalences among different Multi-tape machines and among Multi and Single-tape machines}
\label{s322_Turing}

In the classical framework  $k$-tape Turing machines have the same
computational power of Single-tape Turing machines and given a
Multi-tape Turing Machine $\M_{K}$ it is always possible to define a
Single-tape Turing Machine which is able to fully simulate its
behavior and therefore to completely execute its computations. As
showed for Single-tape Turing machine (see \cite{Sergeyev_Garro}),
the Grossone methodology allows us to give a more accurate
definition of the equivalence among different machines as it
provides the possibility not only to separate different classes of
infinite sets with respect to their cardinalities but also to
measure the number of elements of some of them. With reference to
Multi-tape Turing machines, the Single-tape Turing Machines adopted
for their simulation use a particular kind of tape which is divided
into tracks (multi-track tape). In this way, if the tape has $m$
tracks, the head is able to access (for reading and/or writing) all
the $m$ characters on the tracks during a single operation. This
tape organization leads to a straightforward definition of the
behavior of a Single-tape Turing machine able to completely execute
the computations of a given Multi-tape Turing machine (see Section
\ref{s12_Turing}). However, the so defined Single-tape Turing
machine $\M$, to simulate $t$ computational steps of $\M_{K}$, needs
to execute $O(t^{2})$ transitions ($t^{2}+t$ in the worst case) and
to use an alphabet with
$2^{k}(\left|\Sigma_{1}\right|+1)\prod^{k}_{i=2}(\left|\Sigma_{i}\right|+2)$
symbols (again see Section~\ref{s12_Turing}). By exploiting the
Grossone methodology is is possibile to obtain the following result
that has a higher accuracy with respect to that provided by the
traditional framework.

\begin{theorem}
Let  us consider $\M_{K}=\seq{Q, \Gamma, \bar{b}, \Sigma, q_0, F,
\delta^{(k)}}$,a $k$-tapes, $k\geq2$, Turing machine, where
$\Sigma=\bigcup^{k}_{i=1}\Sigma_{i}$ is given by the union of the
symbols in the k tape alphabets $\Sigma_{1},\dots,\Sigma_{k}$ and
$\Gamma=\Sigma\cup\{\bar{b}\}$. If this machine performs t
computational steps such that
 \beq
  t\leqslant\frac{1}{2}(\sqrt{4\mbox{\ding{172}}+1}-1),
\label{Turing_ks}
 \eeq
then there exists $\M'=\{ Q', \Gamma', \bar{b}, \Sigma', q_0', F', \delta' \}$, an equivalent Single-tape Turing machine with $\left|\Gamma'\right|=2^{k}(\left|\Sigma_{1}\right|+1)\prod^{k}_{i=2}(\left|\Sigma_{i}\right|+2)$, which is able to simulate $\M_{K}$ and can be observed in a sequence.
\end{theorem}

\textit{Proof.} Let us recall that the  definition of $\M'$ requires
for a Single-tape to be divided into $2k$ tracks; $k$ tracks for
storing the characters in the $k$ tapes of $\M_{K}$ and $k$ tracks
for signing through the marker $\downarrow$ the positions of the $k$
heads on the $k$ tapes of $\M_{k}$ (see Section~\ref{s12_Turing}).
The transition function $\delta^{(k)}$ of the $k$-tapes machine is
given by
$\delta^{(k)}(q_{1},{a_{i}}_{1},\dots,{a_{i}}_{k})=(q_{j},{a_{j}}_{1},\dots,{a_{j}}_{k},{z_{j}}_{1},\dots,{z_{j}}_{k})$,
with ${z_{j}}_{1}, \dots,{z_{j}}_{k} \in\{R,L,N\}$; as a consequence
the corresponding transition function $\delta'$ of the Single-tape
machine, for each transition specified by $\delta^{(k)}$ must
individuate the current state and the position of the marker for
each track and then write on the tracks the required symbols, move
the markers and go in another internal state. For each computational
step of $\M_{K}$, $\M'$ must execute a sequence of steps for
covering the portion of tapes between the two most distant markers.
As in each computational step a marker can move at most of one cell
and then two markers can move away each other at most of two cells,
after $t$ steps of $\M_{K}$ the markers can be at most $2t$ cells
distant, thus if $\M_{K}$ executes $t$ steps, $\M'$ executes at
most: $2\sum^{t}_{i=1}i = t^{2}+t$ steps. In order to be observable
in a sequence the number $t^{2}+t$ of steps, performed by $\M'$ to
simulate $t$ steps of $\M_{K}$, must be less than or equal to
\ding{172}. Namely, it should be $t^{2}+t\leqslant$\ding{172}. The
fact that this inequality is satisfied for $
t\leqslant\frac{1}{2}(\sqrt{4\mbox{\ding{172}}+1}-1)$ completes the
proof. \hfill $\Box$

\section{Concluding Remarks}
\label{s4_Turing} In the paper, Single and Multi-tape  Turing
machines have been described and observed through the lens of the
Grossone language  and methodology. This new language, differently
from the traditional one, makes it possible to distinguish among
infinite sequences of different length so enabling a more accurate
description of Single and Multi-tape Turing machines. The
possibility to express the length of an infinite sequence explicitly
gives the possibility to establish more accurate results regarding
the equivalence of machines in comparison with the observations that
can be done   by using the traditional language.

It is worth noting  that the traditional results and those presented
in the paper do not contradict one another. They are just written by
using different mathematical languages having different accuracies.
Both mathematical languages observe and describe the same objects --
Turing machines -- but with different accuracies. As a result, both
traditional and new results are correct with respect to the
mathematical languages used to express them and correspond to
different accuracies of the observation. This fact is one of the
manifestations of the relativity of mathematical results formulated
by using different mathematical languages in the same way as the
usage of a stronger lens in a microscope gives a possibility to
distinguish more objects within an object that seems to be unique
when viewed by a weaker lens.

Specifically, the Grossone language has allowed us to give the
definition of \textit{complete output} of a Turing machine, to
establish when and how the output of a machine can be observed, and
to establish a more accurate relationship between a Multi-tape
Turing machine and a Single-tape one which simulates its
computations. Future research efforts will be geared to apply the
Grossone language and methodology to the description and observation
of new and emerging computational paradigms.



\bibliographystyle{plain}

\end{document}